\documentclass[11pt,epsf]{article}
\usepackage{amsmath}
\usepackage{amsfonts}
\usepackage{amssymb}
\usepackage{graphicx}
\usepackage{color}
\usepackage{comment}
\newcommand {\dfn} {\stackrel{\Delta} {=}}
\newcommand {\exe} {\stackrel{\cdot} {=}}
\newcommand {\lexe} {\stackrel{\cdot} {\le}}

\newcommand {\reals} {{\rm I\!R}}

\newcommand {\bs} {\mbox{\boldmath $s$}}

\newcommand {\bx} {\mbox{\boldmath $x$}}

\newcommand {\bz} {\mbox{\boldmath $z$}}

\newcommand {\bB} {\mbox{\boldmath $B$}}

\newcommand {\bE} {\mbox{\boldmath $E$}}

\newcommand {\bS} {\mbox{\boldmath $S$}}

\newcommand {\bX} {\mbox{\boldmath $X$}}

\newcommand{\calC}{{\cal C}}

\newcommand{\calG}{{\cal G}}

\newcommand{\calN}{{\cal N}}
\newcommand{\calO}{{\cal O}}

\newcommand{\calT}{{\cal T}}

\newcommand{\calV}{{\cal V}}

\allowdisplaybreaks

\begin{document}
\thispagestyle{empty}
\title{Trade-offs Between Weak-Noise Performance and Probability of Anomaly in
Parameter Estimation from Noisy Chaotic Signals}
%\thanks{This research was supported by my wife and kids.}
\author{Neri Merhav
%\thanks{
%Currently on sabbatical leave at HP Laboratories,
%1501 Page Mill Road, MS 3U-4, Palo Alto CA 94304, USA.}
}
\date{}
\maketitle

\begin{center}
The Andrew \& Erna Viterbi Faculty of Electrical Engineering\\
Technion - Israel Institute of Technology \\
Technion City, Haifa 32000, ISRAEL \\
E--mail: {\tt merhav@ee.technion.ac.il}\\
\end{center}
\vspace{1.5\baselineskip}
\setlength{\baselineskip}{1.5\baselineskip}

\begin{abstract}
We consider the problem of parameter estimation,
based on noisy chaotic signals, from the viewpoint of twisted modulation
for waveform communication. In particular, we study communication systems
where the parameter to be estimated is conveyed as the initial condition  
of a chaotic dynamical system of a certain class and we examine its
estimation performance in terms of the expectation of a given convex function
of the estimation error at high SNR, under
the demand that the probability of anomaly is kept small. We derive a
lower bound on the weak-noise estimation error for this class of chaotic
modulators, and argue that it can be outperformed by using the itinerary signal
associated with the chaotic
system instead of the main chaotic output signal.\\

\end{abstract}

\newpage
\section{Introduction}

We revisit the well known problem of twisted modulation and estimation, that
is, conveying the value of a parameter $\theta$ by
$n$ uses of an additive white Gaussian noise (AWGN) channel,
\begin{equation}
y_i=x_i+z_i,~~~~~~~i=1,2,\ldots,n,
\end{equation}
where $x_i$ is the $i$--th component of a channel input vector,
$x^n=(x_1,x_2,\ldots,x_n)=f_n(\theta)$, that depends on
the parameter $\theta$, and that is subjected to a power constraint,
$\|\bx\|^2\le nQ$, $\{z_i\}$ are independent, zero--mean, Gaussian
random variables with variance $\sigma^2$, and $y_i$ is the $i$--th coordinate
of the channel output vector, $y^n=(y_1,y_2,\ldots,y_n)$.
In general, the main interest, in twisted modulation and estimation, is to quantify
how well can
one estimate $\theta$
based on $y^n$ when it is allowed to optimize both the modulator, $f_n$, and the estimator.
In particular, how fast does the estimation error decay as a function of $n$
when the best modulator and estimator are used? In this work, these questions
are addressed in the context of modulators that are based
on a certain class of chaotic dynamical systems, but before we confine
ourselves to those chaotic modulators, we first discuss the twisted modulation
problem in general.

The twisted modulation and estimation problem, which is the
discrete--time counterpart of the
``waveform communication'' problem (according to the terminology of
\cite[Chap.\ 8]{WJ65}),
can be addressed both from the information--theoretic and the
estimation--theoretic viewpoints. In the realm of information theory, this falls within the
framework of joint source--channel coding (see, e.g., \cite{KGT17} and references
therein), where a single source symbol $\theta$ is conveyed by $n$ channel
uses, and it also falls within the formalism of Shannon--Kotel'nikov
mappings (see, e.g., \cite{Floor08}, \cite{Hekland07}, \cite{HFR09},
\cite{KR07} and
references therein). From the estimation--theoretic perspective, any given
modulator gives rise to some parametric family of conditional probability
density functions of
$y^n$ given $\theta$, and then estimation theory provides a rich variety of both Bayesian
and non--Bayesian lower
bounds (depending on whether $\theta$ is deterministic or random)
on the estimation performance (mostly, in terms of the mean square error), as
well as a plethora of useful estimators that
perform well at least for high signal-to-noise ratio (SNR), for example, the maximum likelihood (ML) estimator in the non--Bayesian
case, the maximum
a--posteriori (MAP) estimator in the Bayesian case, estimators based on the
method of moments, and others.

A central well known
problem, that is inherent to all non--linear (twisted) modulators and
receivers, is the {\it threshold effect}
(see, e.g., \cite[Chap.\ 8]{WJ65}). The threshold effect amounts to a sudden
transition between two modes of behavior when the
SNR crosses a certain critical value. For large SNR, namely, in the {\it weak--noise}
regime, the estimation error of the ML estimator behaves
similarly as that of a (locally) linear modulator, and so, it roughly
achieves the Cram\'er--Rao lower bound (CRLB). But beyond a certain noise level, the
estimation performance breaks down completely and rather abruptly. As explained in
\cite[Chap.\ 8]{WJ65}, for a given non--linear modulator, one can identify a
certain {\em anomaly event} (or outage event), whose probability becomes
considerably large as the threshold SNR is crossed downwards.

The rich literature on non--linear modulation and estimation
contains many performance bounds (see, e.g.,
\cite{Burnashev84}, \cite{Burnashev85}, \cite{Cohn70} and many references
therein), with no distinction between weak--noise errors and
anomalous errors, in other words, both types of errors weigh in the
evaluation of the total mean square error (MSE). 
However, in view of the above background, it makes a lot of sense to separate the two kinds of
errors because estimation under anomaly is rather
meaningless. Accordingly, the idea separating the two types of errors appear
already in \cite[pp.\ 661--674]{WJ65}, but not quite in a formal manner.
A more methodological approach in this direction, of separating weak--noise errors
from anomalous errors, appears in \cite[Section IV.A]{KGT17}, where the
problem was posed in terms of designing a communication system, along with a
definition of the anomaly event,
with the objective of minimizing the MSE given
that the anomaly event has not occurred, subject to the requirement
that the probability of anomaly would be less than a given constant.
It was shown in \cite{KGT17}, that the data processing lower bound is
asymptotically attained by a simple modulator, that first quantizes
the parameter $\theta$ and then maps the quantized parameter into a good
digital channel code for the Gaussian channel. The receiver first decodes the
digital message and then maps it back to the corresponding quantized parameter
value. The outage event is then the error event in the digital part and then
the weak-noise MSE is simply the quantization error.
The weak link in this result,
however, is that the data processing lower bound is not quite compatible to
this setting, as it corresponds to a situation where there is no freedom to
allow a definition of an outage event as part of the design.

In a more recent work \cite{me19}, the approach of \cite{KGT17} was refined in two ways: first,
both in the lower bound and in the upper bound, an anomaly
event is allowed, with the freedom to define it, depending on the
communication system itself. In other words, the
lower bound and the upper bound in \cite{me19} refer to the same setting, in contrast to
\cite{KGT17}. Secondly, 
instead of limiting the probability of anomaly to be upper bounded by a small
constant, in \cite{me19} we set the constraint that the probability of anomaly would not
exceed the function $e^{-\alpha n}$ for some
given constant $\alpha > 0$. Under this constraint, the
fastest possible decay rate of the MSE
is studied in \cite{me19},
or more generally, the expectation of
an arbitrary symmetric, convex function of the estimation error is bounded.
More precisely, we derived in \cite{me19} an upper bound and a lower bound, which agree in
the limit of high SNR for a certain range of values of $\alpha$.

In this paper, we study twisted modulators that
are based on a certain class of chaotic dynamical systems, where the parameter
$\theta$ to be conveyed is encoded in the initial condition of the chaotic
sequence that is transmitted into the channel. More specifically, we consider
modulators based on chaotic systems whose
transmitted output sequence is (a scaled and shifted version of)
$s^n=(s_1,s_2,\ldots,s_n)$, that is generated recursively according to
\begin{equation}
\label{ds}
s_{i+1}=q(s_i),~~~~~i=0,1,2,\ldots
\end{equation}
where the initial condition is $s_0=\theta$ and $q:[0,1]\to[0,1]$ is a piece-wise linear function
whose (absolute) slope is larger than unity at all points of continuity of $q$
(i.e., almost everywhere).
The use of chaotic dynamical systems for modulation and estimation has become
very popular in the last three decades, see, e.g., \cite{Chen96},
\cite{CXS99}, \cite{CW98},
\cite{Drake98}, \cite{HM04}, \cite{KK00}, \cite{LSXW06}, \cite{PVLS03}, \cite{PW95}, \cite{Wallinger13},
\cite{WYL99}, \cite{XTCL09}, \cite{YWLZ18}
for a non-exhaustive sample of references, which altogether cover the topic from a
variety of aspects, including upper and lower bounds on Bayesian/non-Bayesian
estimation, numerical aspects, algorithmic efficiency, system optimization, extensions and
applications in Turbo coding, in hybrid coding, in spread spectrum systems,
and in MIMO systems, just to name a few. One of the main reasons for the
popularity of the use chaotic systems for modulation
is the strong sensitivity of the output signal,
$s^n$, to small perturbations in the initial condition, $\theta$, which is a very
desirable property for accurate estimation of $\theta$ at the receiver which observes
a noisy version of $s^n$, or of some memoryless transformation of $s^n$.

The contribution of the present work is that it studies chaotic modulators
systematically within the
framework of waveform communication established in \cite[Chap.\ 8]{WJ65}. In
particular, we examine the above described class of chaotic systems using the
results of \cite{me19} and thereby derive a lower bound on the weak-noise
estimation error subject to the requirement of small probability of anomaly.
It is believed that this framework can be useful for examining other classes
of modulators (chaotic and others) as well.
In contrast to many of the above cited papers on chaotic modulation,
we avoid the use of the Cram\'er-Rao lower bound, which is problematic due to
the fact that, for our class of chaotic modulators, the signal vector is a not continuous function (and a-fortiori, not
differentiable) of the parameter. We use instead
the lower bound of \cite{me19} with the appropriate adjustments required. An
additional benefit of avoiding the Cram\'er-Rao lower bound is that our figure
of merit is more general than the mean square error - it is defined as the
asymptotic exponential decay rate of the expectation of $\rho(\epsilon)$,
where $\epsilon=\hat{\theta}-\theta$ is estimation
error and $\rho(\cdot)$ is an arbitrary convex even cost function with $\rho(0)=0$.

The derivation of this lower bound yields
a limitation on a certain design parameter of the system, depending on
the SNR (details will
follow in the sequel).
Then, we show that by feeding the channel with a secondary
output sequence from the chaotic system, called the itinerary sequence (which is obtained by a certain 
quantization of $\bs$), instead of $\bs$, yields better estimation performance than that of any
modulator from the aforementioned class. In fact, it asymptotically achieves
the lower bound of \cite{me19}, which applies to any non-linear modulator
operating at a given high SNR.
At first glance, this may seem
counter-intuitive because the itinerary sequence seems to convey ``less information'' about
the parameter, but a little thought easily settles this conflict. 

Finally, we discuss the usefulness of the same class of chaotic systems
for the purpose of simulation of random processes, and well as its extension
to systems with long range memory, where eq.\ (\ref{ds}) is replaced by
\begin{equation}
s_{i+1}=q_i(s_0,s_1,\ldots,s_i),~~~~i=0,1,2,\ldots .
\end{equation}
It turns out that for the memoryless Gaussian channel considered here, our
main conclusions continue to apply, in other words, there is no benefit in
using memory of the remote past.

The outline of the remaining part of the paper is as follows.
In Section \ref{sys}, we define the class of chaotic dynamical systems and
their corresponding modulators. In Section \ref{properties}, we explore some
of the properties of these dynamical systems. In Section \ref{lowerbound}, we
derive a lower bound to the weak-noise estimation performance in the
non-outage event. In Section \ref{itinerary}, describe an alternative method
of using the itinerary signal and show that its weak-noise performance can approach the
the lower bound arbitrarily closely. Finally, in Section \ref{longrange}, we
outline the extension to systems with long-range memory and also discuss how
such systems can serve also as optimal process simulators.

\section{The Communication System Setting}
\label{sys}

In this section, we present the class of chaotic systems that we study in this
work, explain how they are used for parameter modulation, and formalize the
objectives of this work.

We begin with the description of the class of dynamical systems that the
modulators are based upon.
Each system in this class is defined by a piece-wise linear map,
parametrized by a positive integer $r\ge 2$ and
a probability vector $P=\{p(0),p(1),\ldots,p(r-1)\}$  
with strictly positive entries,
i.e., $p(x)> 0$ for all $x\in\{0,1,\ldots,r-1\}$ and $\sum_{x=0}^{r-1}p(x)=1$. 
Given $r$ and $P$, we define
\begin{equation}
F(x)=\sum_{x'=0}^{x-1}p(x'),~~~~~~x\in\{0,1,\ldots,r-1\},
\end{equation}
with the convention that the summation over the empty set is defined as zero, namely,
$F(0)=0$. Clearly, $F(r)=1$. Next, define
\begin{equation}
G(x)=\frac{F(x)}{p(x)},~~~~~~~~x\in\{0,1,\ldots,r-1\}.
\end{equation}
Given a real $s\in[0,1]$, let $\phi(s)$ be defined as the unique value of $x\in\{0,1,\ldots,r-1\}$
such that $F(x)\le s < F(x+1)$, where for $x=r-1$, the strong inequality is allowed 
to be weak, namely, $\phi(1)=r-1$. This class of dynamical systems was also
considered in earlier works, such as \cite{WYL99}.

The dynamical system is defined as follows. Given an initial state, $s_0\in[0,1]$,
it generates recursively two sequences, $\bs=(s_1,s_2,\ldots)$ and $\bx=(x_1,x_2,\ldots)$, 
as follows:
For $t=1,2,\ldots$, let
\begin{equation}
\label{recursion}
x_t=\phi(s_{t-1});~~
s_t=\frac{s_{t-1}-F(x_t)}{p(x_t)}.
\end{equation}
We henceforth refer to $\bs$ as the {\em main output sequence}, or as the {\em state
sequence} of the dynamical system, and
to $\bx$ -- as the {\em itinerary sequence} (see also \cite{WYL99}).
Note that for $r=2$ and $p(0)=p(1)=\frac{1}{2}$, this is nothing but the well
known dyadic map, where $s_t=(2s_{t-1})~\mbox{mod}~1$ 
where $u~\mbox{mod}~1$ designates the fractional part of $u$, namely,
$u~\mbox{mod}~1=u-\lfloor u\rfloor$.
Accordingly, $x_1,x_2,\ldots$ are the
bits of the binary representation of $s_0$ as $0.x_1x_2,\ldots$. Likewise,
for a general $r$ and $p(0)=p(1)=\cdot\cdot\cdot=p(r-1)=\frac{1}{r}$ this
is the more general saw-tooth map $s_t=(rs_{t-1})~\mbox{mod}~1$, 
and then
$s_0=0.x_1x_2\ldots$ is the radix-$r$ representation, that is,
$s_0=\sum_{i=1}^\infty x_ir^{-i}$. Accordingly, the system (\ref{recursion})
can be thought of as an extension of these maps that forms a mixed basis
representation of $s_0$. More details will follow in Section \ref{properties}.

The modulators that we consider in this work, are induced by the above defined
dynamical systems as follows. Given a parameter value, $\theta\in[0,1]$, we set the
initial state to be $s_0=\theta$ and then generate the vector $(s_1,s_2,\ldots,s_n)$
according to (\ref{recursion}).\footnote{More generally, we may set $s_0$ to
be some given function of $\theta$.}
Given a prescribed power budget, $Q$,  we then define the channel input signal
according to
\begin{equation}
\label{ut}
u_t=\sqrt{12Q}\left(s_t-\frac{1}{2}\right),~~~~~t=1,2,\ldots,n,
\end{equation}
so that if each $s_t$ is uniformly distributed across the interval $[0,1]$ (as
will be established below),
then $u_t$ is zero-mean and its variance (which is also its power) is exactly
$Q$. The vector $u^n=(u_1,u_2,\ldots,u_n)$ is fed into an additive white
Gaussian noise (AWGN) channel with noise variance $\sigma^2$. The
signal-to-noise ratio (SNR) is then defined as
\begin{equation}
\gamma=\frac{Q}{\sigma^2}.
\end{equation}
Let the channel output vector be denoted by $y^n=(y_1,y_2,\ldots,y_n)$, where
for $i=1,2,\ldots,n$, $y_i=u_i+z_i$, $\{z_i\}$ being realizations of
$\{Z_i\}$,
which are Gaussian, zero-mean,
independently and identically distributed (i.i.d.) random variables with
variance $\sigma^2$. Correspondingly, $\{y_i\}$ are realizations of $\{Y_i\}$
with $Y_i=u_i+Z_i$, $i=1,2,\ldots,n$.
The receiver implements an estimator of $\theta$ based on $Y^n$, namely,
\begin{equation}
\hat{\theta}=\varphi(y^n).
\end{equation}
For every $\theta\in[0,1]$, let $\calO_n(\theta)\subset\reals^n$ designate an event
defined in the
space of noise vectors, $\{\bz\}$, which is henceforth referred to as the {\it
outage event} (or
the {\it anomalous error event}) given $\theta$.
Following the framework of \cite{me19}, 
for a given convex, even error cost function $\rho(\cdot)$, with $\rho(0)=0$,  
our first objective is to
derive a lower bound to
\begin{equation}
\label{obj}
\sup_{\theta\in[0,1]}\bE\left\{\rho(\varphi(Y^n)-\theta)\bigg|\calO_n^{\mbox{\tiny
c}}(\theta)\right\}
\end{equation}
for an arbitrary family of outage events,
subject to the constraint that
\begin{equation}
\label{constraint}
\sup_{\theta\in[0,1]}\mbox{Pr}\{\calO_n(\theta)\}\le e^{-\alpha n},
\end{equation}
where $\alpha> 0$ is prescribed constant, henceforth referred to as the {\em
outage exponent}. Here,
the expectation $\bE\{\cdot\}$ in (\ref{obj}) and the probability
$\mbox{Pr}\{\cdot\}$ in
(\ref{constraint}) are defined with respect
to (w.r.t.) the randomness of the noise vector, $Z_1,Z_2,\ldots,Z_n$. 
We will be interested, first and foremost, in the most relaxed version of the
constraint, where $\sup_{\theta\in[0,1]}\mbox{Pr}\{\calO_n(\theta)\}$ is
merely required to tend to zero, without commitment to a particular
exponential rate. This amounts to the choice $\alpha=\alpha_n\to 0$ in the
limit of large $n$.

\section{Properties of the Dynamical Systems}
\label{properties}

In this section, we explore some properties of the non-linear dynamical
systems defined in Section \ref{sys}. Some of these properties may be of
interest on their own right, and most of them are moreover crucial for our derivation
of the lower bound on the weak-noise estimation performance in Section
\ref{lowerbound}.\\

\noindent
1.~{\em Reconstructing the initial state from the itinerary sequence.}
The first property that is important to mention is the following: given the
infinite itinerary sequence, $\bx=(x_1,x_2,\ldots)$,
one can reconstruct the initial state, $s_0$, using the following expression:
\begin{eqnarray}
\label{s0=w(x)}
s_0&=&w(x_1,x_2,\ldots)\nonumber\\
&\dfn&\sum_{t=1}^\infty F(x_t)\prod_{i=1}^{t-1}p(x_i)\nonumber\\
&\equiv&\sum_{t=1}^\infty G(x_t)\prod_{i=1}^tp(x_i),
\end{eqnarray}
where a product over an empty set of indices ($\prod_{i=1}^0p(x_i)$) is defined as unity.
It is therefore implied that although the itinerary sequence,
$\bx=(x_1,x_2,\ldots)$, is a certain quantized version of the state sequence,
$\bs=(s_1,s_2,\ldots)$, they both bear exactly the same information, as there
is one-to-one correspondence between them via $s_0$: $\bx$ determines $s_0$
via (\ref{s0=w(x)}), which in turn determines $\bs$. But it should be kept in
mind that this one-to-one correspondence holds only for the infinite
sequences. For finite $n$, this is no longer true.
As mentioned earlier, 
the last line of (\ref{s0=w(x)}) can be thought of as a mixed basis representation of $s_0$ in the sense that in the special case where $P$ is the uniform distribution, 
i.e., $p(x)=1/r$ for all $x\in\{0,1,\ldots,r\}$ and then $G(x)=x$, it boils 
down to the radix $r$ representation associated with the saw-tooth map, that
is,
\begin{equation}
s_0=\sum_{t=1}^\infty x_t\cdot r^{-t},
\end{equation}
with $\{x_t\}$ in the role of the digits, i.e., $s_0=0.x_1x_2x_3\ldots$. 

The relation 
(\ref{s0=w(x)}) applies, of course, also to the time-shifted versions
of the sequences, i.e., 
\begin{eqnarray}
s_\tau&=&w(x_{\tau+1},x_{\tau+2},\ldots)\nonumber\\
&=&\sum_{t=\tau+1}^\infty F(x_t)\prod_{i=\tau+1}^{t-1}p(x_i)
\end{eqnarray}
for every positive integer $\tau$.

To see why eq.\ (\ref{s0=w(x)}) holds true,
first observe that if $s_0\in[0,1]$, then so is $s_1$, and then by the same
token, the same is
true for $s_2$, etc., which means that $s_t\in[0,1]$ for all $t\ge 0$.
Inverting the recursion (\ref{recursion}) in $\{s_t\}$,
we have
\begin{equation}
s_{t-1}=F(x_t)+p(x_t)s_t,
\end{equation}
and so,
\begin{eqnarray}
s_0&=&F(x_1)+s_1p(x_1)\nonumber\\
&=&F(x_1)+p(x_1)[F(x_2)+s_2p(x_2)]\nonumber\\
&=&F(x_1)+p(x_1)\{F(x_2)+p(x_2)[F(x_3)+s_3p(x_3)]\}\nonumber\\
&=&F(x_1)+F(x_2)p(x_1)+F(x_3)p(x_1)p(x_2)+s_3p(x_1)p(x_2)p(x_3).
\end{eqnarray}
Likewise, more generally, for every positive integer $\tau$,
\begin{equation}
s_0=\sum_{t=1}^\tau F(x_t)\prod_{i=1}^{\tau-1}p(x_i)+s_\tau\prod_{i=1}^\tau
p(x_i),
\end{equation}
which implies that
\begin{equation}
\label{bounds0}
\sum_{t=1}^\tau F(x_t)\prod_{i=1}^{\tau-1}p(x_i)\le s_0\le \sum_{t=1}^\tau
F(x_t)\prod_{i=1}^{\tau-1}p(x_i)+[\max_xp(x)]^\tau,
\end{equation}
where we have used the fact that $0\le s_\tau\le 1$. It now follows that when
$\tau\to\infty$, 
$s_0$ is sandwiched between two sequences (indexed by $\tau$) that both tend to
$\sum_{t=1}^\infty F(x_t)\prod_{i=1}^{t-1}p(x_i)$.\\

\noindent
2.~{\em The itinerary sequence as a random process.}
Let $S_0$ be a random variable, uniformly distributed across the unit interval, $[0,1]$, and
let $\bS=(S_1,S_2,\ldots)$ and $\bX=(X_1,X_2,\ldots)$ the corresponding 
sequences of random variables generated from $S_0$ according to the recursion (\ref{recursion}).
Observe that the numbers, 
$F(x)$, $x\in\{0,1,\ldots,r-1\}$ can be viewed as endpoints of successive, non-overlapping 
sub-intervals of
lengths $p(x)$, $x\in\{0,1,\ldots,r-1\}$, forming a partition of the unit
interval. It is therefore clear that 
$X_1$ is distributed according to $P$.
Now, given $X_1=x_1$, $S_0$ is clearly uniformly distributed across the interval $[F(x_1),F(x_1+1)]$.
Thus, the equation $S_1=[S_0-F(X_1)]/p(X_1)$ shifts and stretches this uniform distribution, 
such that $S_1$ is again uniform across $[0,1]$, which means that the uniform distribution over
$[0,1]$ is invariant under the recursion. Since $S_1$ is uniform over $[0,1]$ independently of
$X_1$, then $X_2$ is distributed according to 
$P$ for the same reason that $X_1$ was distributed 
according to $P$ for $S_0\sim\mbox{Unif}[0,1]$. Moreover, since $S_1$ is independent of $X_1$,
then $X_2=\phi(S_1)$ is also independent of $X_1$. Likewise, $S_2$ is uniformly distributed
over $[0,1]$, independently of 
$(X_1,X_2)$, and so, $X_3=\phi(S_2)$ is distributed according to $P$, independently of $(X_1,X_2)$, and so on. It follows then that $\bX$ is a sequences of i.i.d.\
random variables, all with distribution $P$. In the sequel, it will be useful
to use the equivalence between the uniform distribution over the initial
state, $S_0$, and the fact that the induced process, $X_1,X_2,\ldots$, is
i.i.d.\ with a marginal distribution $P$, via the one-to-one relationship between
them, as was established in item no.\ 1 above.\\

\noindent
3.~{\em The Lyapunov exponent.}
Since the class of dynamical systems considered here are chaotic in general, 
one of the important characteristics is the Lyapunov exponent, which measures
the local exponential rate of divergence between two state sequences that
begin from two very close initial states (see, e.g., \cite{ER85}).
Assuming that $S_0$ is uniformly distributed over $[0,1]$, so that
$X_1,X_2,\ldots$ are i.i.d.\ with distribution $P$,
the Lyapunov exponent is given by
\begin{eqnarray}
\lambda&=&\lim_{n\to\infty}\frac{1}{n}\bE\left\{\log\bigg|\frac{\partial
s_n}{\partial s_0}\bigg|\right\}\nonumber\\
&=&\lim_{n\to\infty}\frac{1}{n}\bE\left\{\log\bigg|\prod_{i=1}^n\frac{\partial
s_i}{\partial s_{i-1}}\bigg|\right\}\nonumber\\
&=&\lim_{n\to\infty}\frac{1}{n}\bE\left\{\log\prod_{i=1}^n\frac{1}
{p(X_i)}\right\}\nonumber\\
&=&H,
\end{eqnarray}
where $H$ is the entropy the source $P$, i.e.,
\begin{equation}
H=-\sum_{x=0}^{r-1}p(x)\log p(x).
\end{equation}
In words, the Lyapunov exponent coincides with the entropy of the itinerary
process, which is maximized when $P$ is the uniform distribution.\\

\noindent
4.~{\em The length of the signal locus.}
An important factor in the weak-noise performance of a twisted modulation system
is the length of the signal locus \cite[Chap.\ 8]{WJ65} (see also
\cite{me19}): Consider the channel input vector (\ref{ut}), with the
(temporary) notation
$u^n(\theta)=(u_1(\theta),\ldots,u_n(\theta))$
that emphasizes the dependence on the parameter $\theta$.
While $\theta$ exhausts the
interval $[0,1]$, the vector $u^n(\theta)$ draws a curve in the vector space
of dimension $n$.
The length of this curve, in general, is given by
\begin{equation}
\label{Ln}
L_n=\int_0^1\|\dot{u}^n(\theta)\|\mbox{d}\theta,
\end{equation}
where $\dot{u}^n(\theta)=(\dot{u}_1(\theta),\ldots,\dot{u}_n(\theta))$,
$\dot{u}_i(\theta)$ being the derivative of $u_i(\theta)$ w.r.t.\ $\theta$ (provided that it
exists), $i=1,2,\ldots,n$.
In our case, each $u_i(\theta)$ is a piece-wise linear function of $\theta=s_0$
with discontinuities. Therefore, the vector $u^n(\theta)$, as a whole, is also
piece-wise linear and piece-wise continuous in $\theta$ and there are exactly
$r^n-1$ points of discontinuity along the interval $0\le\theta\le 1$, because
as $s_0$ varies from $0$ to $1$, the itinerary vector $x^n$ exhausts all $r^n$
possible values, beginning from
$(0,0,\ldots,0)$ and ending at $(r-1,r-1,\ldots,r-1)$.
Each change in one of the components
of $x^n$ corresponds to one discontinuity point of $s^n$.
The integral in (\ref{Ln}) expresses the sum of
lengths of the continuous pieces of the signal locus. To assess $L_n$ in our
case, consider the following.
Along each sub-interval of continuity, we have
\begin{eqnarray}
\dot{u}_i(\theta)&=&\frac{\mbox{d}u_i(\theta)}{\mbox{d}\theta}\nonumber\\
&=&\frac{\mbox{d}u_i(s_0)}{\mbox{d}s_0}\nonumber\\
&=&\frac{\mbox{d}u_i}{\mbox{d}s_i}\prod_{j=1}^i\frac{\mbox{d}s_j}{\mbox{d}s_{j-1}}\\
&=&\sqrt{12Q}\cdot\prod_{j=1}^i\frac{1}{p(x_j)}
\end{eqnarray}
Thus, $L_n$, is upper bounded as follows:
\begin{eqnarray}
L_n&=&\sqrt{12Q}\int_0^1\mbox{d}s_0\cdot\sqrt{\sum_{i=1}^n\prod_{j=1}^i\frac{1}{p^2(x_j)}}\nonumber\\
&=&\sqrt{12Q}\bE\left\{\sqrt{\sum_{i=1}^n\prod_{j=1}^i\frac{1}{p^2(X_j)}}\right\}\nonumber\\
&\le&\sqrt{12Q}\bE\left\{\sqrt{n\cdot\prod_{j=1}^n\frac{1}{p^2(X_j)}}\right\}\nonumber\\
&=&\sqrt{12Qn}\cdot\sum_{\bx}\prod_{i=1}^np(x_i)\sqrt{\prod_{j=1}^n\frac{1}{p^2(x_j)}}\nonumber\\
&=&\sqrt{12Qn}\cdot r^n,
\end{eqnarray}
where we have use the fact integration along the interval $s_0\in[0,1]$ is
equivalent to expectation w.r.t.\ the randomness of the random variable
$S_0\sim\mbox{unif}[0,1]$, which in turn is equivalent to expectation w.r.t.\
the randomness of $\{X_i\}$ according being an i.i.d.\ process with
probability distribution $P$. The inequality in the above chain amounts to the
fact that the sum of $n$ positive terms cannot exceed $n$ times the largest
term, which is the last one.
Likewise,
\begin{eqnarray}
L_n&=&\sqrt{12Q}\bE\left\{\sqrt{\sum_{i=1}^n\prod_{j=1}^i\frac{1}{p^2(X_j)}}\right\}\nonumber\\
&\ge&\sqrt{12Q}\bE\left\{\sqrt{\prod_{j=1}^n\frac{1}{p^2(X_j)}}\right\}\nonumber\\
&=&\sqrt{12Q}\sum_{\bx}\prod_{i=1}^np(x_i)\sqrt{\prod_{j=1}^n\frac{1}{p^2(x_j)}}\nonumber\\
&=&\sqrt{12Q}r^n,
\end{eqnarray}
which means that $L_n\exe r^n$, independent of $P$. Here, 
the notation $a_n\exe b_n$, for two positive sequences,
$\{a_n\}$ and $\{b_n\}$, means that $\frac{1}{n}\log\frac{a_n}{b_n}\to 0$, as
$n\to\infty$.\\

\noindent
5.~{\em The autocorrelation function and the spectrum of the state process.}
We now calculate the autocorrelation individual function of the stationary state process $\{S_t\}$,
induced by a uniformly distributed initial state, $S_0$,  across the interval $[0,1]$.
In Appendix A, we derive the following expression for $R_S(k)\dfn\bE\{S_0S_k\}$
($k$ -- integer):
\begin{equation}
R_S(k)=\frac{1}{4}+\frac{1}{12}\cdot\left(\sum_{x=0}^{r-1}p^2(x)\right)^{|k|},
\end{equation}
where the first term comes from the DC component of $\{S_t\}$, which is
$\frac{1}{2}$, as each $S_t$ is distributed uniformly over $[0,1]$. The
exponential term is the more interesting term. Using the relation (see
(\ref{ut}),
\begin{equation}
U_t=\sqrt{12Q}\left(S_t-\frac{1}{2}\right),
\end{equation}
it is clear that the autocorrelation function of $\{U_t\}$, is given by
\begin{equation}
R_U(k)=\bE\{U_0U_k\}=Q\cdot\left(\sum_{x=0}^{r-1}p^2(x)\right)^{|k|}.
\end{equation}
For the sake of brevity, let us denote 
\begin{equation}
\bar{p}=\bE\{p(X)\}=\sum_{x=0}^{r-1}p^2(x),
\end{equation} 
and then
the power spectrum is
\begin{equation}
S_U(e^{j\theta})=\frac{Q(1-\bar{p}^2)}{(1-\bar{p}e^{-j\theta})(1-\bar{p}e^{j\theta})},~~~~~~~j\dfn\sqrt{-1}.
\end{equation}
The channel output spectrum is then
\begin{eqnarray}
S_Y(e^{j\theta})&=&\frac{Q(1-\bar{p}^2)}{(1-\bar{p}e^{-j\theta})(1-\bar{p}e^{j\theta})}+\sigma^2\nonumber\\
&=&\frac{A(1-\tau e^{-j\theta})(1-\tau e^{j\theta})}{(1-\bar{p}e^{-j\theta})(1-\bar{p}e^{j\theta})}
\end{eqnarray}
where $A$ and $\tau$ are readily found to be given by:
\begin{equation}
\tau=\frac{2\sigma^2\bar{p}}{\sigma^2(1+\bar{p}^2)+Q(1-\bar{p}^2)+\sqrt{[\sigma^2(1+\bar{p}^2)+Q(1-\bar{p}^2)]^2-4\sigma^2\bar{p}^2}}<1
\end{equation}
and
\begin{equation}
\label{Adef}
A=\frac{1}{2}\left[\sigma^2(1+\bar{p}^2)+Q(1-\bar{p}^2)+\sqrt{[\sigma^2(1+\bar{p}^2)+Q(1-\bar{p}^2)]^2-4\sigma^2\bar{p}^2}\right].
\end{equation}

\noindent
{\em 6.~Ergodic properties.}
Since $R_U(k)$ tends to zero as $k\to\infty$, then so does
$\frac{1}{2k+1}\sum_{j=-k}^kR_U(j)$, and then it follows by Slutsky's theorem
\cite[p.\ 432]{Papoulis91} that the process $\{U_t\}$
is mean-ergodic, that is,
\begin{equation}
\lim_{n\to\infty}\bE\left\{\left(\frac{1}{n}\sum_{t=1}^nU_t-\bE\{U_1\}\right)^2\right\}=
\lim_{n\to\infty}\bE\left\{\left(\frac{1}{n}\sum_{t=1}^nU_t-\right)^2\right\}=0.
\end{equation}
Using the inequalities (\ref{bounds0}), it is straightforward to check
(see Appendix B for the details)
that $\{U_t\}$ is also autocovariance-ergodic \cite[p.\ 437]{Papoulis91},
namely, 
\begin{equation}
\lim_{n\to\infty}\bE\left\{\left[\frac{1}{n}\sum_{t=1}^nU_tU_{t+k}-R_U(k)\right]^2\right\}=0
\end{equation}
for every fixed non-negative integer $k$. This is done by applying Slutsky's theorem
to establish mean-ergodicity of the process $V_t=U_tU_{t+k}$. It then follows
by Chebychev's inequality that for every $\epsilon> 0$,
\begin{equation}
\lim_{n\to\infty}\mbox{Pr}\bigg\{\bigg|\frac{1}{n}\sum_{t=1}^nU_tU_{t+k}-R_U(k)\bigg|>\epsilon\bigg\}=0.
\end{equation}
Since $\{u_t\}$ are completely determined by $s_0$,
then the randomness of $\{U_t\}$ fully stems from the randomness of the
uniformly distributed initial state, $S_0$, and then it follows that the Lebesgue
measure of the set 
\begin{equation}
\left\{s_0:~\bigg|\frac{1}{n}\sum_{t=1}^nu_tu_{t+k}-R_U(k)\bigg|\le\epsilon,~k=0,1\right\},
\end{equation}
tends to unity as $n\to\infty$. Since $\{u_t\}$ are piece-wise continuous
functions of $s_0$, then so is $\frac{1}{n}\sum_{t=1}^nu_tu_{t+k}$, which
implies that the above set is the union of disjoint sub-intervals of
$[0,1]$ whose total Lebesgue measure (length) tends to unity.\\

\noindent
7.~{\em An upper bound on the mutual information between the channel input and output.}
The induced normalized mutual information associated with the channel is then:
\begin{eqnarray}
\label{C1def}
C_0&=&\lim_{n\to\infty}\frac{h(Y^n)-h(Y^n|X^n)}{n}\nonumber\\
&\le&\frac{1}{4\pi}\int_{-\pi}^{\pi}\ln[2\pi e
S_Y(e^{j\theta})]\mbox{d}\theta-\frac{1}{2}\ln(2\pi e\sigma^2)\nonumber\\
&=&\frac{1}{4\pi}\int_{-\pi}^{\pi}\ln[
S_Y(e^{j\theta})]\mbox{d}\theta-\frac{1}{2}\ln \sigma^2\nonumber\\
&=&\frac{1}{2}\ln\left(\frac{A}{\sigma^2}\right)\nonumber\\
&\dfn&C_1.
\end{eqnarray}
Note that $A$ can also be written as
\begin{equation}
A=\frac{1}{2}\left[\sigma^2+Q+\bar{p}^2(\sigma^2-Q)+\sqrt{(\sigma^4+Q^2)(1-\bar{p}^2)^2+2\sigma^2Q(1-\bar{p}^4)}\right],
\end{equation}
and so,
\begin{equation}
\frac{A}{\sigma^2}=\frac{1}{2}\left[1+\gamma+\bar{p}^2(1-\gamma)+\sqrt{(1+\gamma^2)(1-\bar{p}^2)^2+2\gamma(1-\bar{p}^4)}\right],
\end{equation}
which makes it clear that, at least when $\gamma>1$, $A/\sigma^2$ and hence
also $C_1$, is a decreasing function of
$\bar{p}$. 

\section{A Lower Bound on the Weak-Noise Estimation Performance}
\label{lowerbound}

In this section, we derive a lower bound on the weak-noise estimation
performance of twisted modulators that are based on the class of chaotic
dynamical systems that were defined in Section \ref{sys}. We do this on
the basis of the generic lower bound of \cite{me19}, but with the appropriate adjustments 
that are particular to this class of modulators, using the results of Section
\ref{properties}. The resulting bound is then tighter than the generic bound of
\cite{me19}. For the sake of completeness, we begin by presenting the generic lower bound of
\cite{me19} as background, and the first step to this end, is to provide a few definitions
and to list the assumptions therein.

Given a positive real $\alpha$, which designates the outage exponent
(\ref{constraint}), let
$w(\alpha)$ be the unique solution
$w$ to
the equation
\begin{equation}
\label{walpha}
w-\ln(1+w)=2\alpha.
\end{equation}
Let us also define
\begin{equation}
R(\alpha,\gamma)=\frac{1}{2}\ln\frac{\gamma}{1+w(\alpha)}.
\end{equation}
The following assumptions are imposed in \cite{me19}.\\

\noindent
A.1 For any given constant $c >
0$, the error cost function $\rho$ obeys 
$\rho(e^{-nc})\exe e^{-n\zeta(c)}$, where $\zeta(\cdot)$
is some continuous function
with the property that $c> 0$ implies $\zeta(c) > 0$.\\

\noindent
A.2 Denoting $M_n=e^{nR
(\alpha,\gamma)}$,
consider the partition of the unit interval into $M_n$
non--overlapping sub--intervals, each of size $1/M_n$. Then,
for all sufficiently large $n$, the number, $M_n^{\mbox{\tiny c}}$,
of sub--intervals in which $s^n(\theta)$ is continuous, obeys
$M_n^{\mbox{\tiny c}}\exe e^{nR
(\alpha,\gamma)}$.\\

\noindent
A.3 If, for a certain vector $\bz\in \calO_n(\theta)$, one of the
components vanishes, then upon replacing this component by any non--zero
number, the resulting vector remains in $\calO_n(\theta)$.
In addition, it is assumed that as noise variance tends to zero, the covering
radius of $\calO_n^{\mbox{\tiny c}}(\theta)$ tends to zero as well.\\

Denoting
\begin{equation}
E(\alpha,\gamma)=\zeta[R(\alpha,\gamma)],
\end{equation}
the following lower bound is derived
in \cite[eq.\ (21)]{me19} under assumptions A1-A3:
\begin{equation}
\label{lbme19}
\sup_{0\le\theta\le
1}\bE\left\{\rho(\varphi(Y^n)-\theta)\bigg|\calO_n^{\mbox{\tiny c}}(\theta)\right\}\ge
\exp\{-n[E(\alpha,\gamma)+o(\gamma)]\},
\end{equation}
where $o(\gamma)$ designates a function that tends to zero as
$\gamma\to\infty$.
In \cite{me19}, this lower
bound is derived in two steps. In the first step, it is shown that
for every positive integer $M$,
\begin{equation}
\label{1ststep}
\sup_{0\le\theta\le
1}\bE\left\{\rho(\varphi(Y^n)-\theta)\bigg|\calO_n^{\mbox{\tiny
c}}(\theta)\right\}\ge
2\rho\left(\frac{1}{2M}\right)\cdot\left[Q\left(\frac{L_n}{2\sigma
M}\right)-e^{-\alpha n}\right],
\end{equation}
where $L_n$ is the length of the locus of the signal fed into the channel (for
our modulators, $\{u^n(\theta),~0\le\theta\le
1\}$), as
described in Section \ref{properties}. In the second step, it is proved that to
meet the outage constraint, $L_n$ must be upper bounded according to
\begin{equation}
L_n\le \sigma\cdot\exp\{n[R(\alpha,\gamma)+o(\gamma)]\}.
\end{equation}
By substituting this upper bound instead of $L_n$ in the right-hand side of
(\ref{1ststep}), we further lower bound $\sup_{0\le\theta\le
1}\bE\left\{\rho(\varphi(Y^n)-\theta)\bigg|\calO_n^{\mbox{\tiny
c}}(\theta)\right\}$ (as $Q(\cdot)$ is a decreasing function), and then by selecting $M$ to be of the exponential order
of $\exp\{n[R(\alpha,\gamma)\}$, the lower bound (\ref{lbme19}) is obtained.
The upper bound on $L_n$ is obtained by a `tube-packing' argument, in the
spirit of the one in \cite[pp.\ 669--674]{WJ65}, which is a version of the
sphere-packing argument of coding for the Gaussian channel (see, e.g., \cite[p.\ 265]{CT06}), but adapted to parameter
modulation and estimation. According to this argument, the tube
generated by the union of spheres of radius $\sqrt{n\sigma^2}$ around all
vectors along the
curve $\{u^n(\theta),~0\le\theta\le 1\}$, must be packed within the sphere of
vectors $\{y^n\}$ that are obtained as $y^n=u^n+z^n$, where $u^n$ is within a
sphere of radius $\sqrt{nQ}$ and $z^n$ is within a sphere of radius
$\sqrt{n\sigma^2}$ and is orthogonal to $u^n$, namely, a sphere of radius
$\sqrt{n(Q+\sigma^2)}$. Another way to look at this sphere is as the set
of $y^n$-vectors defined as follows: Let $Y=U+Z$, where $U\sim\calN(0,Q)$ and
$Z\sim\calN(0,\sigma^2)$ be independent random variables, and let $f_Y(y)$ be
the (Gaussian) probability density function (pdf) obtained by the convolution
between $f_U(u)$ - the pdf of $U$, and $f_Z(z)$ - the pdf of $Z$. Now, let
$h(Y)$ be the differential entropy of $Y$. Then, the
sphere of radius $\sqrt{n(Q+\sigma^2)}$ can be thought of as the
typical set 
\begin{equation}
\calT_n(Y)=\left\{y^n:~\frac{1}{n}\sum_{i=1}^n\ln f_Y(y_i)\ge
-h(Y)-\epsilon\right\},
\end{equation}
and then let $\epsilon$ tends to zero (after letting $n$ grow without bound).
Indeed,
\begin{equation}
1\ge\int_{\calT_n(Y)}\mbox{d}y^n\cdot\prod_{i=1}^nf_Y(y_i)\ge
\int_{\calT_n(Y)}\mbox{d}y^n\cdot
e^{-n[h(Y)+\epsilon]}=\mbox{Vol}\{\calT_n(Y)\}\cdot e^{-n[h(Y)+\epsilon]},
\end{equation}
and so,
\begin{equation}
\mbox{Vol}\{\calT_n(Y)\}\le
e^{n[h(Y)+\epsilon]}=\exp\left\{n\left[\frac{1}{2}\ln(2\pi
e\sigma^2)+\epsilon\right]\right\}.
\end{equation}
We now modify this lower bound, along with its derivation in \cite{me19}, by particularizing it to the class of
chaotic modulators
considered here. The difference is in the upper bound to $L_n$. Note that the
vector $u^n(\theta)$ must lie within the $n$-dimensional hyper-cube
$[-\sqrt{12Q},+\sqrt{12Q}]^n$. A typical noise vector, $z^n$, is essentially
orthogonal (uncorrelated) to $u^n(\theta)$, and so, a typical channel output
vector, $y^n=u^n(\theta)+z^n$, is the sum of a vector in
$[-\sqrt{12Q},+\sqrt{12Q}]^n$ and an orthogonal vector in a sphere of radius
$\sqrt{n\sigma^2}$. To assess the volume of typical channel output vectors,
we proceed along the same line as before:
Consider a random variable $Y=U+Z$, where $U$ is uniformly distributed across
the interval $[-\sqrt{12Q},+\sqrt{12Q}]$ and $Z\sim\calN(0,\sigma^2)$
independent of $U$. Accordingly, let $f_Y(y)$ be the
pdf of $Y$, which is now the convolution between the uniform pdf of $U$
over $[-\sqrt{12Q},+\sqrt{12Q}]$ and the Gaussian pdf of $Z$. Let $h(Y)$
denote the differential entropy associated with $f_Y$, and consider 
again the set $\calT_n(Y)$ but with the new definitions of $f_Y$ and $h(Y)$.
Then, here too, $\mbox{Vol}\{\calT_n(Y)\}$ is essentially upper bounded by
$e^{nh(Y)}=e^{nh(U+Z)}$. According to a tube-packing bound similar to that of \cite{WJ65}
and \cite{me19}, we now have (ignoring the $\epsilon$-term):
\begin{equation}
L_n\lexe \frac{e^{nh(Y)}}{(2\pi
e\sigma^2[1+w(\alpha)])^{(n-1)/2}}\exe\exe\sqrt{2\pi
e}\cdot\sigma\cdot\exp\left\{n\left[h(Y)-\frac{1}{2}\ln(2\pi
e\sigma^2[1+w(\alpha)])\right]\right\}.
\end{equation}
For large $\gamma$, the convolution of the rectangular $f_U$ with the Gaussian
$f_Z$ is very close to the original $f_U$, and so, $h(Y)=
h(U)+o(\sigma^2)=\frac{1}{2}\ln[12Q(1+\Delta(\gamma))]$, where
$\Delta(\gamma)\to 0$ as $\gamma\to \infty$.
Thus,
\begin{eqnarray}
L_n&\le&\sqrt{2\pi
e}\cdot\sigma\cdot\exp\left\{\frac{n}{2}\ln\left(\frac{12Q(1+\Delta(\gamma))}{2\pi
e\sigma^2[1+w(\alpha)]}\right)\right\}\nonumber\\
&=&\sqrt{2\pi
e}\cdot\sigma\cdot\exp\left\{n\left(\frac{1}{2}\ln\gamma-\frac{1}{2}\ln\left(\frac{2\pi
e}{12}\right)+\frac{1}{2}\ln[1+\Delta(\gamma)]-\frac{1}{2}\ln[1+w(\alpha)]
\right)\right\}\nonumber\\
&\dfn&L_n^*.
\end{eqnarray}
Consider now the case of very high SNR ($\gamma\gg 1$) and $\alpha=\alpha_n\to 0$, which
means that although we wish to keep the probability of anomaly small, we do
not insist on an exponential decay of this probability. In this case, we can
neglect the last two terms at the exponent and only the first two terms
remain. The term $\frac{1}{2}\ln\gamma$ approximates the channel capacity,
$C=\frac{1}{2}\ln(1+\gamma)$ for large $\gamma$. The constant term $\frac{1}{2}\ln(2\pi
e/12)\approx 0.1765$, which we henceforth denote by $\mu$, is the loss due to channel input shaping mismatch. 

Next, recall that in Section \ref{properties}, we have seen that for the
chaotic modulators considered here, $L_n\ge \sqrt{12Q}r^n$, independent of $P$. Together with the upper bound
$L_n\le L_n^*\exe \exp\{n[C-\mu+o(\gamma)]\}$, it follows that the anomaly
probability constraint imposes a limitation on the alphabet size, $r$. Specifically, neglecting the
$o(\gamma)$ term at the exponent of $L_n^*$, we see that $r$ must not exceed
$e^{C-\mu}$. Upon selecting $r <
e^{C-\mu}$, we have $r^n \ll e^{n(C-\mu)}$ for large $n$, which means, among
other things, that the number of discontinuity points of $u^n(\theta)$, as
function of $\theta$, which is $r^n-1$, is exponentially smaller than $e^{n(C-\mu)}$, a fact
that in turn allows us to
select, similarly as in \cite{me19}, $M\exe L_n^*/(2\pi s)$ (with $s$ being
an arbitrary positive
constant) in (\ref{1ststep}) and be assured that Assumption A.2
is satisfied. The resulting lower bound would then be
\begin{equation}
\sup_{0\le\theta\le
1}\bE\left\{\rho(\phi(Y^n)-\theta)\bigg|\calO_n^{\mbox{\tiny
c}}(\theta)\right\}\ge \rho\left(\frac{\sigma
s}{L_n^*}\right)\left[Q(s)-e^{-n\alpha_n}\right]\exe e^{-n\zeta(C-\mu)},
\end{equation}
and so, the loss compared to the generic lower bound of $\cite{me19}$ is at
least in the shaping loss reduction by $\mu$.

{\em Discussion.} It should be pointed out the shaping mismatch is not the
only factor that causes loss in performance to our class of chaotic modulators
considered here. 
The reason is that
there is an additional inherent limitation on the channel input signals
generated by these chaotic modulators, and this is the fact that they possess
memory. More precisely, owing to the ergodic
properties of the dynamical system (as discussed in Section \ref{properties}),
for most values of
$\theta\in[0,1]$ (in the Lebesgue measure sense), $u^n(\theta)$ possesses an
empirical autocorrelation function that is close to
$R_U(k)=Q\cdot\bar{p}^{|k|}$. This further reduces the volume of the object in
which the tube of noisy signals must be packed. A detailed analysis of this 
volume reveals that it is upper bounded by the exponential order of $A^{n/2}$,
where $A$ is defined as in eq.\ (\ref{Adef}), and so, another upper bound to
$L_n^*$ is of the exponential order of $e^{nC_1}$, where $C_1$ is defined as
in (\ref{C1def}). This alternative upper bound, however, ignores the fact that
$u^n(\theta)$ is limited to lie in the hyper-cube
$[-\sqrt{12Q},+\sqrt{12Q}]^n$. Combining the two limitations on $u^n(\theta)$
in a joint manner is not a trivial task, but of course, one can always take the better
between the two individual bounds. Anyway, we will not delve here into this analysis
because the first lower bound above suffices on its
own right to make the
point that the performance of the chaotic modulators cannot approach the
generic lower bound, not even in the exponential order. Nonetheless, there is
an alternative way to use the these modulators so that the lower bound of
\cite{me19} would be approached arbitrarily closely in the exponential order.
This is the subject of the next section.

\section{Feeding the Channel by the Itinerary Signal}
\label{itinerary}

So far, we have considered our chaotic modulators as devices that generate and feed the channel with
$u^n$, which is a scaled and shifted version of the state sequence,
$s^n$, as was also done in most of the earlier works
on twisted modulation using chaotic dynamical systems. However, our dynamical
system generates also an additional output, that is, the itinerary sequence
$x^n=(x_1,\ldots,x_n)$. In this section, we consider the option of feeding the
channel with $x^n$, for the purpose estimating $\theta$. More precisely, we
allow some mapping $c:\{0,1,\ldots,r-1\}\to\calV$, where $\calV$ is a set of
$r$ real numbers $\{c(x),~x\in\{0,1,\ldots,r-1\}\}$ that designate legitimate channel inputs that meet the power
constraint, $\sum_{x=0}^{r-1}p(x)[c(x)]^2\le Q$, and we transmit
$v^n=c(x^n)=(c(x_1),\ldots,c(x_n))$.

At first glance, it may not seem like a good idea to use $x^n$ or any function
of it, such as $c(x^n)$, as an alternative to $s^n$, because each $x_i$ is a quantized
version of $s_{i-1}$, and so, it seems to convey less information about
$\theta$. Somewhat surprisingly, it turns out that with the correct choices, it works
better than with $s^n$, or $u^n$. Note that while it is true that the itinerary vector is a quantized version
of the corresponding state vector, it turns out that in the limit of large $n$, the difference
between them vanish in terms of the information they convey on $\theta$.
Indeed, recall that the infinite sequence $\bx=(x_1,x_2,\ldots)$ fully determines
$\theta$, which in turn fully determines $\bs=(s_1,s_2,\ldots)$. Also,
in contrast to the transmission of $u^n$, if one uses $v^n=c(x^n)$ as the channel input vector,
there is no longer a limitation on $r$ as we had before, and there
are therefore many degrees of
freedom. In fact, $r$ can even be countably infinite. As it turns out, by using $x^n$ with the correct
choices of $r$, $P$, $c(\cdot)$, and a certain map $s_0=\psi(\theta)$, where
$\psi:[0,1]\to[0,1]$ (see footnote no.\ 1), we can approach the generic lower bound of
\cite{me19} arbitrarily closely at least for the case $\alpha\to 0$.

Consider the following construction: Let $\epsilon> 0$ be
arbitrarily small and select $R=C-\epsilon$, where
$C=\frac{1}{2}\ln(1+\gamma)$ is the capacity of the Gaussian channel. Let
$M=e^{nR}$ and consider the grid of $M$ points, $\calG=\{1/(2M),3/(2M),5/(2M),\ldots, (2M-1)/(2M)\}$.
For each grid point $\hat{\theta}_i=(2i-1)/2M$, $i=1,2,\ldots,M$, select independently at random 
a number, denoted $\psi(\hat{\theta}_i)$, under the uniform distribution over
the interval $[0,1]$. Reveal these independent random selections to both
transmitter and receiver. With probability one, these $M$ random numbers are
all distinct, and then the inverse map $\psi^{-1}$ is well defined. 
Let $r$ be an arbitrarily large odd number and let 
\begin{equation}
c(x)=\delta\cdot\left(x-\frac{r-1}{2}\right),~~~~x=0,1,\ldots,r-1,
\end{equation}
where $\delta> 0$ is an arbitrarily small number such that
$r\delta\gg\sqrt{Q}$. Finally, let $P$ be defined
as follows:
\begin{equation}
p(x)=\left\{\begin{array}{ll}
\int_{-\infty}^{-(r/2-1)\delta}g(a)\mbox{d}a & x=0\\
\int_{\delta(x-r/2)}^{\delta(x-r/2+1)}g(a)\mbox{d}a & x\in\{1,2,\ldots,r-2\}\\
\int_{(r/2-1)\delta}^\infty g(a)\mbox{d}a & x=r-1\end{array}\right.
\end{equation}
where $g(a)=(2\pi Q)^{-1/2}e^{-a^2/(2Q)}$.

Our modulation scheme works as follows. Given a parameter value
$\theta\in[0,1]$, consider its quantization to the nearest grid point
$\hat{\theta}_i\in\calG$ and then define $s[i]=\psi(\hat{\theta}_i)$ as the
initial state. Let $x^n[i]$ be the corresponding itinerary sequence generated by the system,
starting at $s[i]$. Transmit $c(x^n[i])$ over the channel, and at the receiver side
apply a ML channel decoder for the code $\calC=\{c(x^n[i]),~i=1,2,\ldots,M\}$.
Let $\hat{i}$ be the index of the decoded message. Then, the estimated
parameter is $\hat{\theta}=\psi^{-1}[s(\hat{i})]$.

To see why this scheme nearly achieves the lower bound, is conceptually simple. The quantization error in $\theta$
cannot exceed $\frac{1}{2M}= \frac{1}{2}e^{-n(C-\epsilon)}$ whose cost is
$\rho\left(\frac{1}{2}e^{-n(C-\epsilon)}\right)\exe e^{-n\zeta(C-\epsilon)}$.
Since $\{\psi(\hat{\theta}_i),~i=1,\ldots,M\}$ are
independent uniformly distributed random variables, their corresponding
itinerary vectors $\{c(x^n[i]),~i=1,\ldots,M\}$ are independent i.i.d.\ random vectors that form a
codebook for the Gaussian channel, where the random coding distribution is a
finely quantized version of the capacity-achieving pdf, $g(a)$, and hence it
is nearly capacity-achieving for small $\delta$ and large $r$. Since $R < C$,
the error probability of the decoding, which is viewed as the probability of anomaly, is
arbitrarily small. In the event of correct decoding, $\hat{i}=i$, the correct
$\hat{\theta}_i$ is reconstructed and error cost remains
$e^{-n\zeta(C-\epsilon)}$, namely, the quantization error cost only.

\section{Extension to Dynamical Systems with Long-Range Memory}
\label{longrange}

The class of chaotic dynamical systems that we considered so far can be
naturally extended to possess longer range memory of the past and the lower
bound derived in Section \ref{lowerbound} will continue to apply.
Specifically, given an integer $r\ge 2$ and a sequence of conditional probability
distributions,
$\{p_t(x_t|x_1,\ldots,x_{t-1}),~x_1,x_2,\ldots,x_t\in\{0,1,\ldots,r-1\}\}$,
$t=1,2,\ldots$, defining a probability law of a process $P$, let us define
\begin{equation}
F_t(x|x_1,\ldots,x_{t-1})=\sum_{x'=0}^{x-1}p_t(x'|x_1,\ldots,x_{t-1}),
\end{equation}
and let
$\phi_t(s|x_1,\ldots,x_{t-1})$ 
be defined as the unique value of $x\in\{0,1,\ldots,r-1\}$
such that
\begin{equation}
F_t(x|x_1,\ldots,x_{t-1})\le s < F_t(x+1|x_1,\ldots,x_{t-1}),
\end{equation}
where for $x=r-1$, the strong inequality is allowed
to be weak, namely, $\phi(1|x_1,\ldots,x_{t-1})=r-1$.
The dynamical system is now defined by the following recursion:
\begin{equation}
\label{recursion1}
x_t=\phi_t(s_{t-1}|x_1,\ldots,x_{t-1});~~
s_t=\frac{s_{t-1}-F_t(x_t|x_1,\ldots,x_{t-1})}{p_t(x_t|x_1,\ldots,x_{t-1})};~~t=1,2,\ldots
\end{equation}
Eq.\ (\ref{s0=w(x)}) extends to this case reads as follows:
\begin{eqnarray}
s_0&=&\sum_{t=1}^\infty
F_t(x_t|x_1,\ldots,x_{t-1})\prod_{j=1}^{t-1}p_j(x_j|x_1,\ldots,x_{j-1})\nonumber\\
&=&\sum_{t=1}^\infty
F_t(x_t|x_1,\ldots,x_{t-1})p_{t-1}(x_1,\ldots,x_{t-1}).
\end{eqnarray}
If the initial state is a random variable, $S_0$, uniformly distributed over
the interval $[0,1]$, then the induced itinerary sequence, $X_1,X_2,\ldots$,
is a random process whose probability law is governed by the given sequence of
conditional probability distributions,
$\{p_t(x_t|x_1,\ldots,x_{t-1}),~x_1,x_2,\ldots,x_t\in\{0,1,\ldots,r-1\}\}$,
$t=1,2,\ldots$. 
Here, the Lyapunov exponent is given by the entropy rate,
\begin{equation}
\lambda=\bar{H}=\lim_{n\to\infty}\frac{H(X_1,\ldots,X_n)}{n},
\end{equation}
provided that the limit exists.
The length of the signal locus continues to be upper bounded by $\sqrt{12Qn}\cdot r^n$
and lower bounded by $\sqrt{12Q}r^n$. If the itinerary signal is used for transmission
over a general channel (that may not be necessarily memoryless),
it naturally suggests to select $P$ to be capacity-achieving channel input process.

Finally, as a side remark, we
point out that
the mapping between $S_0$ and $\bX=(X_1,X_2,\ldots)$ can serve as a simulator
for synthesizing a prescribed random process $P$ from a sequence of purely
random bits \cite{HV93}.
Consider the binary symmetric source of random bits,
$\bB=(B_1,B_2,\ldots)$ that define the uniformly distributed random variable,
\begin{equation}
\label{binaryrepresentation}
S_0=\sum_{i=1}^\infty B_i2^{-i},
\end{equation}
which in turn is mapped to the sequence of random variables,
$\bX=(X_1,X_2,\ldots)$ governed by the process law $P$. Viewing this
as a mapping from
$\bB$ to $\bX$, it can be considered a simulator of a random
process with an optimal average conversion rate \cite{HV93} of
$\bar{H}$ bits per symbol, where $\bar{H}$ is the entropy rate, assuming it exists. To see why this is true, assume first that
$P$ is a dyadic source in the sense that all conditional distributions,
$p_t(x_t|x_1,\ldots,x_{t-1})$, are integer powers of $\frac{1}{2}$, and
consider the recursion:
\begin{eqnarray}
S_t&=&\frac{S_{t-1}-F_t(X_t|X_1,\ldots,X_{t-1})}{p_t(X_t|X_1,\ldots,X_{t-1})}\nonumber\\
&=&\exp_2\{\log[1/p_t(X_t|X_1,\ldots,X_{t-1})]\}\cdot[S_{t-1}-F_t(X_t|X_1,\ldots,X_{t-1})].
\end{eqnarray}
In terms of the binary representation of $S_{t-1}$,
the subtraction of $F_t(X_t|X_1,\ldots,X_{t-1})$ amounts to a certain manipulation of the bits of
$S_{t-1}$ whereas
the multiplication by $\exp_2\{\log[1/p_t(X_t|X_1,\ldots,X_{t-1})]\}$ means a shift of
$\log[1/p_t(X_t|X_1,\ldots,X_{t-1})]$ bits to
the left. After the shift, the left-most $\log[1/p_t(X_t|X_1,\ldots,X_{t-1})]$
bits are discarded and $S_t$ then depends on the tail, starting
from $(\log[1/p_t(X_t|X_1,\ldots,X_{t-1})]+1)$-th bit and onward. Thus, to generate $X_t$,
$\log[1/p_t(X_t|X_1,\ldots,X_{t-1})]$ bits were utilized, and so, on the average, we have
used $\bE\{\log[1/p_t(X_t|X_1,\ldots,X_{t-1})]\}=H(X_t|X_1,\ldots,X_{t-1})$ bits per symbol. 
Averaging over $t=1,2,\ldots,n$ yields $H(X_1,\ldots,X_n)/n$, which tends to
$\bar{H}$ is the limit exists.
For a non-dyadic source,
$\log[1/p_t(X_t|X_1,\ldots,X_{t-1})]$ are not all integers, but then to reduce the effect
of rounding errors, one considers the effect of $n\gg 1$ successive
iterations, with a total shift of
$\sum_{t=1}^n\log[1/p_t(X_t|X_1,\ldots,X_{t-1})]=\log[1/p_t(X_1,\ldots,X_t)]$,
and then applies the usual considerations well known
from the theory of variable-length lossless source coding. The advantage of
this simulator is
that it is relatively easy to implement and that it
generates a process with the exact distribution,
as opposed to fixed-rate schemes with the approximate distribution.

\begin{comment}
The inverse mapping can serve for random bit extraction.
On the face of it, it has an infinite delay as $S_t$ depends on the whole tail
of $\bX$ beginning from $X_{t+1}$. However, the delay can be limited by taking
advantage of the fact that for large $t$, the contributions of the $\{X_t\}$
become negligible. Consider the approximation obtained by truncation of the
infinite summation into a finite one, i.e.,
\begin{equation}
\hat{S}_0=\sum_{t=1}^kF(X_t)\prod_{i=1}^{t-1}p(X_t).
\end{equation}
Then, the approximation error is bounded by
\begin{eqnarray}
0&\le&S_0-\hat{S}_0\nonumber\\
&=&\sum_{t=k+1}^\infty F(X_{t})\prod_{i=1}^{t-1}p(X_{t})\nonumber\\
&\le&\sum_{t=k+1}^\infty \prod_{i=1}^{t-1}p(X_{t})\nonumber\\
&\le&\sum_{t=k+1}^\infty p_{\max}^{t-1}\nonumber\\
&=&\frac{p_{\max}^k}{1-p_{\max}},
\end{eqnarray}
and so, $(X_1,\ldots,X_k)$ is guaranteed to dictate
$(B_1,B_2,\ldots,B_{\ell})$, provided
that
\begin{equation}
2^{-\ell}\ge\frac{p_{\max}^k}{1-p_{\max}},
\end{equation}
or equivalently,
\begin{equation}
\ell\le k\cdot\log_2\left(\frac{1}{p_{\max}}\right)+\log_2(1-p_{\max}).
\end{equation}
\end{comment}

\section*{Appendix A}
\renewcommand{\theequation}{A.\arabic{equation}}
    \setcounter{equation}{0}

\subsection*{Derivation of $R_S(k)$}

To derive $R_S(k)$, we use the relationship between the state at a given time and the
itineraries at all later times.
For a given integer $k\ge 0$, we define
\begin{eqnarray}
R_S(k)&\dfn&\bE\{S_0S_k\}\nonumber\\
&=&\bE\left\{\sum_{t=1}^\infty
F(X_t)\prod_{i=1}^{t-1}p(X_t)\cdot\sum_{\ell=1}^\infty
F(X_{k+\ell})\prod_{j=1}^{\ell-1}p(X_{k+j})\right\}\nonumber\\
&=&\sum_{t=1}^\infty\sum_{\ell=1}^\infty A(t,\ell),
\end{eqnarray}
where
\begin{equation}
A(t,\ell)=\bE\left\{F(X_t)F(X_{k+\ell})\prod_{i=1}^{t-1}p(X_t)\cdot\prod_{j=1}^{\ell-1}p(X_{k+j})\right\}.
\end{equation}
We denote
\begin{eqnarray}
\bar{p}&=&\bE\{p(X)\}=\sum_{x=0}^{r-1}p^2(x)\\
\bar{F}&=&\bE\{F(X)\}=\sum_{x=0}^{r-1}p(x)F(x)\\
\overline{p^2}&=&\bE\{p^2(X)\}=\sum_{x=0}^{r-1}p^3(x)\\
\overline{F^2}&=&\bE\{F^2(X)\}=\sum_{x=0}^{r-1}p(x)F^2(x)\\
\overline{pF}&=&\bE\{p(X)F(X)\}=\sum_{x=0}^{r-1}p^2(x)F(x).
\end{eqnarray}
For calculating $A(t,\ell)$, we distinguish between several cases.\\

\noindent
1.\ {\em The case $k+1>t$.} In this case, there is no overlap between the
segments $X_1^t$ and $X_{k+1}^{k+\ell}$, and so, they are independent.
Thus,
\begin{equation}
A(t,\ell)=(\bar{F})^2(\bar{p})^{t+\ell-2}.
\end{equation}

\noindent
2.\ {\em The case $k+1=t$.} In this case, there is an overlap of one sample,
$X_t=X_{k+1}$, and so,
\begin{equation}
A(t,\ell)=(\bar{p})^{t-1}\cdot\overline{pF}\cdot(\bar{p})^{\ell-2}\cdot\bar{F}=(\bar{p})^{t+\ell-3}\cdot\bar{F}\cdot\overline{pF}.
\end{equation}

\noindent
3.\ {\em The case $k+1<t$.} This case should be subdivided into three
secondary
sub-cases:\\

\noindent
3a.\ {\em The sub-case $t>k+\ell$.}
In this case, the segment $X_1^t$ fully covers the segment $X_{k+1}^{k+\ell}$,
and then
\begin{equation}
A(t,\ell)=(\bar{p})^k(\overline{p^2})^{\ell-1}\cdot\overline{pF}\cdot(\bar{p})^{t-k-\ell-1}\cdot\bar{F}=
(\bar{p})^{t-\ell-1}\cdot\overline{pF}\cdot\bar{F}\cdot(\overline{p^2})^{\ell-1}.
\end{equation}

\noindent
3b.\ {\em The sub-case $t=k+\ell$.}
Similar to 3a, except that $X_t\equiv X_{k+\ell}$.
\begin{equation}
A(t,\ell)=(\bar{p})^k\cdot(\overline{p^2})^{\ell-1}\cdot\overline{F^2}.
\end{equation}

\noindent
3c.\ {\em The sub-case $t<k+\ell$.}
Here the segments $X_1^t$ and $X_{k+1}^{k+\ell}$ have only a partial overlap
and none of them is a sub-string of the other.
\begin{equation}
A(t,\ell)=(\bar{p})^k\cdot(\overline{p^2})^{t-k-1}\cdot\overline{pF}\cdot(\bar{p})^{k+\ell-t-1}\cdot\bar{F}=
(\bar{p})^{2k+\ell-t-1}\cdot(\overline{p^2})^{t-k-1}\cdot\bar{F}\cdot\overline{pF}.
\end{equation}

Putting everything together, we get:
\begin{eqnarray}
R_S(k)&=&\sum_{t=1}^\infty\sum_{\ell=1}^\infty A(t,\ell)\nonumber\\
&=&\sum_{t=1}^k\sum_{\ell=1}^\infty A(t,\ell)+\sum_{\ell=1}^\infty
A(k+1,\ell)+\sum_{t=k+2}^\infty\left[\sum_{\ell=1}^{t-k-1}A(t,\ell)+A(t,t-k)+\sum_{\ell=t-k+1}^\infty
A(t,\ell)\right]\nonumber\\
&=&\sum_{t=1}^k\sum_{\ell=1}^\infty
(\bar{F})^2(\bar{p})^{t+\ell-2}+\sum_{\ell=1}^\infty
(\bar{p})^{t+\ell-3}\cdot\bar{F}\cdot\overline{pF}+\nonumber\\
&
&\sum_{t=k+2}^\infty\bigg[\sum_{\ell=1}^{t-k-1}(\bar{p})^{t-\ell-1}\cdot\overline{pF}\cdot\bar{F}\cdot(\overline{p^2})^{\ell-1}+
(\bar{p})^k\cdot(\overline{p^2})^{t-k-1}\cdot\overline{F^2}+\nonumber\\
& &\sum_{\ell=t-k+1}^\infty
(\bar{p})^{2k+\ell-t-1}\cdot(\overline{p^2})^{t-k-1}\cdot\bar{F}\cdot\overline{pF}.
\end{eqnarray}
It is easy to verify that all five terms are proportional to the geometric
series $\{\bar{p}^k\}$,
except for the first term, which also has a constant component (independent
of $k$), given by
\begin{eqnarray}
\left[\frac{\bar{F}}{1-\bar{p}}\right]^2&=&\left[\frac{\sum_{x=0}^{r-1}p(x)F(x)}{1-\sum_{x=0}^{r-1}p^2(x)}\right]^2\nonumber\\
&=&\left[\frac{\sum_{x=0}^{r-1}p(x)\sum_{x'=0}^{x-1}p(x')}{\sum_{x=0}^{r-1}\sum_{x'=0}^{r-1}p(x)p(x')-\sum_{x=0}^{r-1}p^2(x)}\right]^2\nonumber\\
&=&\left[\frac{\sum_{(x,x'):~x'<x}p(x)p(x')}{\sum_{(x,x'):~x'\ne
x}p(x)p(x')}\right]^2\nonumber\\
&=&\left(\frac{1}{2}\right)^2
=\frac{1}{4},
\end{eqnarray}
accounting for the DC component
of $\{S_t\}$, which is $\frac{1}{2}$. In other words, $R_S(k)$ is of the form
\begin{equation}
R_S(k)=\frac{1}{4}+q\cdot(\bar{p})^k,
\end{equation}
where $q$ is a constant. The constant $q$ is easily found to be
$q=\frac{1}{12}$ by using the simple fact that
$R_S(0)=\bE\{S_0^2\}=\frac{1}{4}+q$ must be equal to $\frac{1}{3}$ since $S_0$
is uniformly
distributed over $[0,1]$.
It follows then that
\begin{equation}
R_S(k)=\frac{1}{4}+\frac{(\bar{p})^k}{12}.
\end{equation}

\section*{Appendix B}
\renewcommand{\theequation}{B.\arabic{equation}}
    \setcounter{equation}{0}
\subsection*{Autocovariance-Ergodicity of $\{U_t\}$}

Since $\{U_t\}$ is just a scaled and shifted version of $\{S_t\}$ (see
(\ref{ut}), it is
enough to show that the latter has a similar property. We recall the
inequalities (\ref{bounds0}) and denote
\begin{equation}
\hat{S}_0\dfn\sum_{t=1}^\tau F(X_t)\prod_{i=1}^{\tau-1}p(X_i)\le S_0\le
\hat{S_0}+\pi^\tau,
\end{equation}
where $\pi=\max_xp(x)$. Likewise,
\begin{equation}
\hat{S}_k\dfn\sum_{t=k+1}^{\tau+k} F(X_t)\prod_{i=k+1}^{\tau+k-1}p(X_i)\le S_k\le
\hat{S_k}+\pi^\tau.
\end{equation}
Now, let $V_t=U_tU_{t+k}$. If we show that the auto-covariance function of 
$\{V_t\}$ tends to zero, then by Slutsky's theorem, it would be mean-ergodic,
which would then imply that $\{U_t\}$ is autocovariance-ergodic.
Now, let $m\ge k+\tau+1$.
\begin{eqnarray}
\bE\{V_tV_{t+m}\}&=&\bE\{S_0S_kS_mS_{m+k}\}\nonumber\\
&\le&\bE\{[\hat{S}_0+\pi^\tau][\hat{S}_k+\pi^\tau]S_mS_{m+k}\}\nonumber\\
&\le&\bE\{\hat{S}_0\hat{S}_kS_mS_{m_k}\}+2\pi^\tau+\pi^{2\tau}\nonumber\\
&=&\bE\{\hat{S}_0\hat{S}_k\}\bE\{S_mS_{m+k}\}+2\pi^\tau+\pi^{2\tau}\nonumber\\
&\le&\bE\{S_0S_k\}\bE\{S_mS_{m+k}\}+2\pi^\tau+\pi^{2\tau}\nonumber\\
&=&[R_S(k)]^2+2\pi^\tau+\pi^{2\tau}\nonumber\\
&=&\bE\{V_t\}\cdot\bE\{V_{t+m}\}+2\pi^\tau+\pi^{2\tau},
\end{eqnarray}
where the second inequality is due to the fact than $\hat{S}_0\hat{S}_k$
depends on $X_1^{\tau+k}$ whereas $S_mS_{m+k}$ depends on
$X_{m+1},X_{m+2},\ldots$, which are independent.
By choosing $\tau=m-k-1$, we have
\begin{equation}
R_V(m)=\mbox{Cov}\{V_t,V_{t+m}\}=\bE\{V_tV_{t+m}\}-\bE\{V_t\}\cdot\bE\{V_{t+m}\}\le
2\pi^{m-k-1}+\pi^{2(m-k-1)},
\end{equation}
which tends to zero as $m\to\infty$ for every fixed $k$. Likewise,
\begin{eqnarray}
\bE\{V_tV_{t+m}\}
&\ge&\bE\{\hat{S}_0\hat{S}_kS_mS_{m+k}\}\nonumber\\
&=&\bE\{\hat{S}_0\hat{S}_k\}\bE\{S_mS_{m+k}\}\nonumber\\
&\ge&\bE\{(S_0-\pi^\tau)(S_k-\pi^\tau)\}R_S(k)\nonumber\\
&=&[R_S(k)]^2-2\pi^\tau-\pi^{2\tau}\nonumber\\
&=&\bE\{V_t\}\cdot\bE\{V_{t+m}\}-2\pi^\tau-\pi^{2\tau},
\end{eqnarray}
and again, we take $\tau=m-k-1$. Thus, $|\mbox{Cov}\{V_t,V_{t+m}\}|\le
2\pi^{m-k-1}+\pi^{2(m-k-1)}\to 0$ as $m\to\infty$.

\end{document}